# New Insights into Rental Housing Markets across the United States: Web Scraping and Analyzing Craigslist Rental Listings

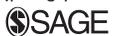

## Geoff Boeing[1] and Paul Waddell[1]

## Abstract

Current sources of data on rental housing—such as the census or commercial databases that focus on large apartment complexes—do not reflect recent market activity or the full scope of the US rental market. To address this gap, we collected, cleaned, analyzed, mapped, and visualized eleven million Craigslist rental housing listings. The data reveal fine-grained spatial and temporal patterns within and across metropolitan housing markets in the United States. We find that some metropolitan areas have only single-digit percentages of listings below fair market rent. Nontraditional sources of volunteered geographic information offer planners real-time, local-scale estimates of rent and housing characteristics currently lacking in alternative sources, such as census data.

## Introduction

It would be difficult to overstate the importance of the rental housing market in the United States, despite longstanding cultural attitudes and policy frameworks encouraging homeownership (Schwartz 2010; Belsky 2013). The share of US households that rent their housing grew from 31 percent in 2004 to 35 percent in 2012, accounting for a total of forty-three million households by 2013 (Joint Center for Housing Studies 2013). Yet a large portion of this rental market activity takes place between private parties leaving minimal and inconsistent data trails. Commercial data sources typically only cover large apartment complexes, and census rental data are limited by their inability to provide reliable current estimates at the local scale or information about unit characteristics. Today, much of the rental listing activity that once occurred in the classified section of local newspapers has moved online to web sites specializing in housing advertisements. The Craigslist web site has become the dominant information exchange in this market and its users generate millions of rental listings each month, yet minimal research has been done to date to explore and understand the rental housing market represented by Craigslist.

To address this knowledge gap in rental housing markets, we collected, cleaned, analyzed, mapped, and visualized eleven million Craigslist US rental listings. The data reveal new insights into spatial patterns in metropolitan housing markets across the United States, and provide much richer detail at much finer scales than other publicly available data sources. New York and San Francisco unsurprisingly have the first and third highest rent per square foot, and North Dakota comes in second, reflecting its recent oil industry boom and housing shortage. We assess affordability by calculating rent burdens and proportions of listings below the US Department of Housing and Urban Development (HUD) fair market rents (FMRs) for 58 metropolitan areas. Although 37 percent of the Craigslist listings in these metropolitan areas are below the corresponding HUD FMR, surprisingly some metropolitan areas like New York and Boston are only in the single-digit percentages. We discover that Craigslist median rents are reasonably comparable to HUD estimates on average, but crucially offer planners more up-to-date data, including unit characteristics, from neighborhood to national scales.

The objectives and motivation for this study are twofold. The first is to present several trends in this underexplored data set and their implications for the housing market. It is the most comprehensive data set currently available to examine the US rental housing market. The second is to share with housing scholars and practitioners a powerful emerging data

**Corresponding Author:**
Geoff Boeing, University of California, Berkeley, 228 Wurster Hall #1850, Berkeley, CA 94720-1850, USA.
Email: gboeing@berkeley.edu

science methodology for collecting and investigating urban data. These methods of data collection, cleaning, and analysis address a growing need for planners to embrace unconventional and emerging tools to explore the vast array of decentralized user-generated data now flowing through cities. The recent explosion in big data and data science has been centered in the fields of computer science, statistics, and physics (O'Neil and Schutt 2013). Planners must understand these tools to help ground—in urban theory and empirical research—the growing urban big data literature being generated in these other fields that have increasingly turned their attention to cities (cf. Bettencourt and West 2010; Bettencourt 2013; Pollock 2016). Yet most importantly, real-time Craigslist data in particular fill a pressing need for planners to measure local-scale rental markets—which evolve quicker than 5-year census rolling averages and data release delays—to understand local conditions, advocate for realistic FMRs, and proactively address emerging affordability challenges.

We begin by providing a brief background on the rental housing market, Craigslist's growing role in it, and the pressing need to study cities through nontraditional sources of big data. Next we explain our methodology for collecting this unique data set, cleaning it, validating it, and analyzing it. Then we present our findings and discuss the practical implications of these housing insights—and urban big data generally—for planners. We conclude with a discussion of the generalizability of our methodology, and the prospects and challenges of big data for planning practitioners and urban scholars.

## Rental Housing Markets and Big Data

Despite the importance of the US rental housing market—particularly in the face of a critical shortage of affordable housing in many cities (Garde 2015)—there are no comprehensive data sources capturing its full scope. Most data used by housing planners come from two sources. The first source is associations of apartment managers and brokers that focus on large apartment complexes, through companies such as CoreLogic, CoStar, Reis, and CBRE. These commercially maintained data sources are valuable, but provide insufficient information about significant segments of the rental market, including garage apartments, condominiums and houses for rent, small self-managed apartment buildings, and granny flats—what Wegmann and Chapple (2012, 35) call "an amateur-operated rental market with some informal characteristics."

The second source is the Census Bureau's American Community Survey (ACS), an invaluable resource for social scientists studying small-scale demographic variation. However, it represents a very small sample of households and can produce inaccurate data (Macdonald 2006; Spielman, Folch, and Nagle 2014). While the annual metropolitan-scale ACS data are useful for broad snapshots of rents, the ACS provides tract-level data only as a five-year rolling average.

Planners thus struggle to acquire up-to-date rental data at the local scale. Further, the ACS rental data provide little information about units. For instance, the median rent for a tract does not reveal what a family of four needs to pay to rent a three-bedroom unit. Practitioners and urban scholars who want to monitor and gain insights into trends across the full spectrum of this market have been unable to do so effectively using existing data sources (Wegmann and Chapple 2013).

Housing rental data from Craigslist address the preceding challenges, yielding millions of observations at fine spatial and temporal scales. Until 10 years ago, available rental units were primarily listed in the classified section of local newspapers. Today they are primarily advertised on web sites like Craigslist, which has become the foremost venue for US rental housing listings (Hau 2006). Craigslist was founded in San Francisco in 1995 by Craig Newmark as an online classified advertisements service (Craigslist 2015). Today, it is the 11th most visited web site in the United States (Alexa 2015) and holds a near monopoly in the online rental listings space (Brown 2014). Between 2000 and 2007, Craigslist took a $5-billion bite out of newspapers' ad revenue (Seamans and Zhu 2014).

Kroft and Pope (2014) found that Craigslist precipitated a 10 percent reduction in average metropolitan rental vacancy rates and increased market efficiency by lowering search costs, thus reducing the average time for units to lease by three weeks. Few researchers have studied Craigslist rental listings directly, but those who have usually do so in the context of landlord discrimination and the Fair Housing Act (e.g., Kurth 2007; Decker 2010; Oliveri 2010; Hanson and Hawley 2011). Mallach (2010), however, hand-tabulated 105 Craigslist listings to estimate the median rent in Phoenix. Wegmann and Chapple (2013) used a small sample of 338 Craigslist listings to study the prevalence of secondary dwelling units in the San Francisco Bay Area. Finally, Feng (2014) web-scraped 6,000 Craigslist listings to study Seattle's housing market.

These listings are a type of Volunteered Geographic Information (VGI), defined as content that is both user-generated and geolocated. VGI is one of the most important and fastest-growing sources of geospatial big data (Jiang and Thill 2015). "Big data," though it varies in interpretation, is more than just a buzzword—it is a type of data that is meaningfully different from traditional and necessarily smaller-scale data (Mayer-Schönberger and Cukier 2013; Kitchin and McArdle 2016). Laney (2001) provides the classic definition of big data, characterized by the three $V$s: volume, variety, and velocity. A fourth $V$, veracity, is sometimes added to signify *volume*'s ability to overcome traditional challenges with messiness and quality. As Jiang and Thill (2015, 1) describe it, "Small data are mainly sampled (e.g., census or statistical data), while big data are automatically harvested . . . from a large population of users." These massive data sets can represent very large samples at incredibly fine spatial and temporal scales, and have significant implications for urban planning and research.

The "smart cities" paradigm promotes harnessing big data for richer understanding, prediction, and planning of cities, though not without controversy (Townsend 2013; Ching and Ferreira 2015; Goodspeed 2015). Big data are starting to have a paradigm-shifting impact on social science research, and data from Internet-based interactions have the potential to reshape our understanding of collective human dynamics (Watts 2007; Batty et al. 2012). Housing markets are ripe for such exploration. Rae (2015) recently looked at eight hundred thousand user-generated housing searches on the British site Rightmove to study the geography of submarkets. However, to date there has been minimal research on large-scale VGI rental listings or the substantial housing market represented by Craigslist.

## Methodology

To narrow this knowledge gap and better understand this market, we collected eleven million rental listings from Craigslist across the United States between May and July 2014. We developed tools to clean the data, extract useful elements, organize them, and analyze them to investigate spatial and temporal patterns—including affordability—in the rental housing market. Throughout, it is important to remember that Craigslist listings provide advertised rents—not final negotiated rents in legal contracts. Metropolitan markets, neighborhoods, and individuals vary in levels of Internet access or technical savvy to list and search for housing online as a function of wealth, race, employment, education, language, social ties, rurality, and other sociodemographic traits (Mossberger et al. 2012). Some rental markets, such as New York's, are dominated by brokers (Gordon 2006). Planners must consider these critical issues in any application of big data or VGI. Nevertheless, Craigslist presents an invaluable data source for housing research.

### Web Scraping

Data are usually transferred over the Internet by means of some formal, structured data set easily processed by a computer. However, the Internet is awash in unstructured and semistructured data never made available as a formal data set: many web pages contain text content that is human-readable but not easily machine-readable. Web scraping bridges this gap and opens up a new world of data to researchers by automatically extracting structured data sets from human-readable content (Mitchell 2015). A *web scraper* accesses web pages, finds specified data elements on the page, extracts them, transforms them if necessary, and finally saves these data as a structured data set. This process essentially mimics how a web browser operates by accessing web pages and saving them to a computer's hard drive cache. In our case, we simply use the contents of this cache for our subsequent analysis after cleaning and organizing the extracted data. A web scraper automates the otherwise cumbersome process of manually collecting data from many web pages and assembling structured data sets out of messy, unstructured text strewn across thousands or even millions of individual pages.

Discussions of web scraping often raise questions of legality and fair use. There are three relevant considerations here: copyright, trespassing, and archives. First, a federal district court decided that it is not a violation of copyright to scrape publicly available data such as Craigslist listings (Craigslist Inc. v. 3Taps Inc. 2013). Moreover, research is a noncommercial fair use that neither repackages nor relists the data. Second, Craigslist *has* previously sued a company—3Taps Inc., who scraped their data for competitive commercial purposes—but only after first sending them a cease-and-desist letter and blocking their IP addresses (ibid.; Splichal 2015). The judge ruled that 3Taps, in effect, trespassed on Craigslist's servers specifically by ignoring the cease-and-desist and using a proxy to circumvent the IP address restrictions that plainly forbid them from accessing the servers (Goldman 2013; Wolfe 2015). Terms of use are subject to change and should be consulted before proceeding on any such project. Third, other organizations such as the Internet Archive (http://archive.org/) scrape and snapshot Craigslist's web pages along with millions of other web sites. Researchers can collect rental listings from these snapshots instead of from Craigslist directly, though they may be less detailed.

We built a web scraper to collect rental listings from the Craigslist web site, using the Python programming language and the *scrapy* web scraping framework (Scrapy Community 2015). First, our web scraper visits a publicly available Craigslist web page that contains rental listings. Next, it receives HTML data back from the web server. This HTML defines the content of web pages (Reid 2015). Then, the scraper extracts the useful data elements from the HTML using the XPath query language (Kay 2008). Finally, our scraper saves these data to a structured data set on a hard drive. We created a process to run the web scraper once each night, configured to collect every Craigslist rental listing that had been posted during the previous day and was still online. During our data gathering, we collected eleven million Craigslist rental listings across 415 regions (i.e., Craigslist's geographic subdomains). This data set covers every rental listing in every Craigslist US subdomain between mid-May and mid-July 2014 (if a listing was posted and taken down on the same day, our scraper did not collect it).

In this study we use the term *region* to refer to these Craigslist subdomains, which can correspond to metropolitan areas, counties, or states depending on the region in question. Craigslist geographies are not always a perfect match for census geographies (e.g., a unit in southern New Hampshire might be listed in either Craigslist's New Hampshire or Boston regions), but the vast majority of listings are far from these gray-area boundaries and the geographies do generally correspond well. For comparability, we used census Combined Statistical Areas when they better matched the Craigslist geography (e.g., in the San Francisco

**Table 1.** Descriptive Statistics for the Data Set at Successive Stages of Processing.

| Descriptive Statistic | Original Data Set | Unique Data Set | Thorough Data Set | Filtered Data Set | Geolocated Data Set |
|---|---|---|---|---|---|
| Count of regions | 415 | 415 | 415 | 415 | 415 |
| Count of listings | 10,958,372 | 5,480,435 | 2,971,362 | 2,947,761 | 1,456,338 |
| Median rent | $1,295 | $1,246 | $1,145 | $1,145 | $1,115 |
| Median ft$^2$ | 1,000 | 982 | 982 | 982 | 960 |
| Median rent/ft$^2$ | $1.09 | $1.11 | $1.11 | $1.11 | $1.10 |
| Mean rent/ft$^2$ | $59.26 | $98.91 | $98.92 | $1.39 | $1.36 |
| Rent/ft$^2$ IQR | 0.89 | 0.89 | 0.89 | 0.88 | 0.83 |
| Rent/ft$^2$ SD | 93,261.96 | 125,082.27 | 125,085.11 | 0.86 | 0.79 |
| Mean bedrooms | 2.13 | 2.08 | 2.06 | 2.05 | 2.04 |

Note: The *original* data set contains the complete original set of listings. The *unique* data set retains one listing per unique ID. The *thorough* data set retains unique listings that contain rent and square-foot data. The *filtered* data set retains thorough listings with reasonable values for rent, square footage, and rent per square foot. The *geolocated* data set retains listings from the filtered data set that contain latitude and longitude. The mean rent per square foot and rent per square foot standard deviation drop sharply between the thorough and filtered data sets, while robust statistics—such as the interquartile range (IQR) and medians—are virtually unchanged. SD = standard deviation.

Bay Area) or conflated Craigslist regions to better match Metropolitan Statistical Areas (MSAs) (e.g., combining Craigslist's Los Angeles and Orange County regions to match the corresponding census MSA).

### Data Cleaning

As is common when collecting VGI, our raw data were very messy. Individual people created these listings through generally free-form text entry, so the rental data we retrieved from Craigslist required substantial filtering and cleaning. We will henceforth refer to the initial, complete, and uncleaned data set as the *original data set*. Descriptive statistics for these data sets and processing steps are summarized in Table 1 and more details are in the appendix.

The first step was the identification and flagging of duplicate listings. Craigslist allows users to resubmit a listing multiple times (retaining the same listing ID), after a couple of days' interval, to restore it to the top of the search results and improve its visibility. Thus, we considered a listing to be a duplicate if its ID appeared more than once in the data set. We will henceforth refer to the set of unique listings as the *unique data set*. Next, we retained only those unique listings with rent and square footage data as the *thorough data set*. Rent and square footage cannot contain negative values: the thorough data set's rent, square footage, and rent per square foot means are much greater than their medians since the distribution is strongly positively skewed by outliers, such as some rents in the billions of dollars. Such values are clearly typos (e.g., billion-dollar rents), spam (e.g., $1 listings linking to an external web site), or other forms of garbage data (e.g., houses listed for sale in the rentals section). Accordingly, we filtered the thorough data set to retain only those listings that had reasonable values for rent, square footage, and rent per square foot.

To define a "reasonable" range, we took the values at the 0.2 percentile and the 99.8 percentile nationwide for each of these three fields as minima and maxima to minimize truncation and provide sensible ranges (Table 2). Thus, a reasonable value is one in the middle 99.6 percent of each variable's distribution. We used these percentiles (rather than the second or third standard deviations above/below the mean) because they provide more realistic value ranges and nationwide criteria give us clear comparability across metropolitan markets. A rent of $10,000 or $12 per square foot is not unheard of in expensive markets like New York or San Francisco, and $189 rent or $0.10 per square foot is plausible for certain properties in the least expensive markets. The range of reasonable square footage values also corresponds to a range from very small studios to large detached homes. We will refer to this set of listings filtered by reasonable values as the *filtered data set*. Lastly, we retained only those rows with latitude and longitude data from the filtered data set as the *geolocated data set*. Listing creators (optionally) assign latitude and longitude by dropping a pin onto an OpenStreetMap interactive web map to explicitly indicate the location of the rental unit. This avoids many of the problems associated with geocoding addresses (e.g., Cayo and Talbot 2003; Zandbergen 2008), yet accuracy depends on the user placing the pin in the correct location.

**Table 2.** Data Filtering Values.

| Variable | Minimum Reasonable Value (0.2 percentile) | Maximum Reasonable Value (99.8 percentile) |
|---|---|---|
| Rent | $189 | $10,287 |
| Ft$^2$ | 220 | 5,200 |
| Rent/ft$^2$ | $0.10 | $12.63 |

### Data Analysis

We analyzed the Craigslist data by region to assess several housing market characteristics, including distributions of rents, square footage, and rents per square foot. To investigate rental affordability patterns, we merged our data set with HUD's 2014 FMR estimates and the 2014 ACS 1-year estimates of median household income and resident population. We used all the listings in the filtered data set for each region that either (1) corresponds to one of the 50 most populous MSAs or (2) is among the 50 regions with the most total listings posted. This sample comprises 58 metropolitan areas and 78 percent of the listings in our filtered data set, and most importantly allows us to compare the Craigslist data to HUD data consistently and with appropriate spatial and population extents.

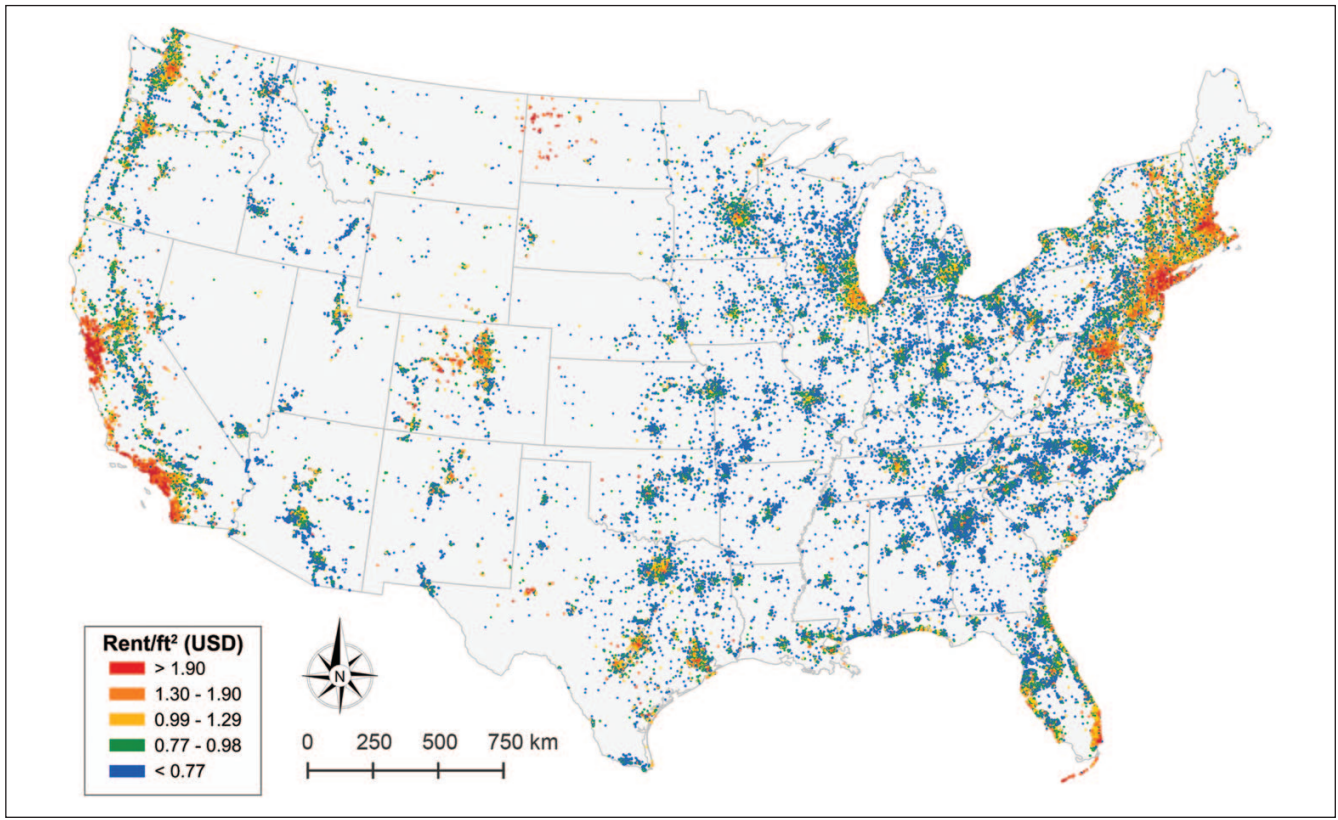

**Figure 1.** Map of the 1.5 million rental listings in the contiguous United States in our geolocated data set.[1]

FMRs are established for policy purposes and generally correspond to 40th percentile rents, "the dollar amount below which 40 percent of the standard-quality rental housing units are rented" (USHUD 2007, p. 1). FMR "areas" generally correspond to metropolitan areas but HUD uses a more complicated formula to determine percentiles and spatial boundaries in certain circumstances (ibid.). For each Craigslist region in the sample, we calculated the rent proportion of income (i.e., the ratio of Craigslist median rent to median monthly household income) and an estimate of how many square feet can be rented in each region for the nationwide median rent (calculated by dividing nationwide median rent by regional median rent per square foot). Then we calculated the proportion of listings in the filtered data set at or below the HUD FMR, per region and number of bedrooms. Finally, we mapped our geolocated data set with a GIS to visualize spatial patterns within and between regions.

## Findings

### National Spatial Patterns

Across the entire filtered data set, the median rent is $1,145, the median square footage is 982, the median rent per square foot is $1.11, and both the mean and median number of bedrooms are approximately 2. The map in Figure 1 depicts 1.5 million rental listings in the contiguous United States in our geolocated data set. Rents per square foot are represented in nationwide quintiles. This map reveals spatial patterns that generally conform to our expectations for the US housing market: large cities on both coasts have higher rents. The map clearly depicts large swaths of high rents per square foot throughout the Boston–Washington corridor and along the coast of California. Other smaller hotspots exist along the coast of southern Florida and in the metropolitan areas of large, affluent cities like Chicago, Denver, and Seattle.

The interior areas of the United States have a sprinkling of less-expensive data points punctuated by middle-quintile clusters around major cities and regional centers. The small towns in the Rocky Mountains west of Denver appear as mini-clusters of expensive rents, because of the significant luxury housing markets in resort towns like Vail and Aspen (Dowall 1981; Lutz 2014). Rental listings in North Dakota generally have extremely high rents per square foot, reflecting recent oil income and unmet demand for housing in oil-producing areas (Holeywell 2011; Brown 2013). In fact, North Dakota's listings have the second-highest median rent per square foot of any region in the entire data set, as the time frame of our data collection preceded the collapse of oil prices in 2015, which put substantial downward pressure on rents in North Dakota (Scheyder 2015). Overall, New York, North Dakota, San Francisco, Boston, and Santa Barbara are the most expensive regions (Figure 2). The other usual suspects from Southern California, Hawaii (cf. Boeing 2016), and the Eastern Seaboard also pepper this list. In contrast, the lowest-priced regions in the data set are small towns across the country that happen to have their own Craigslist subdomains.

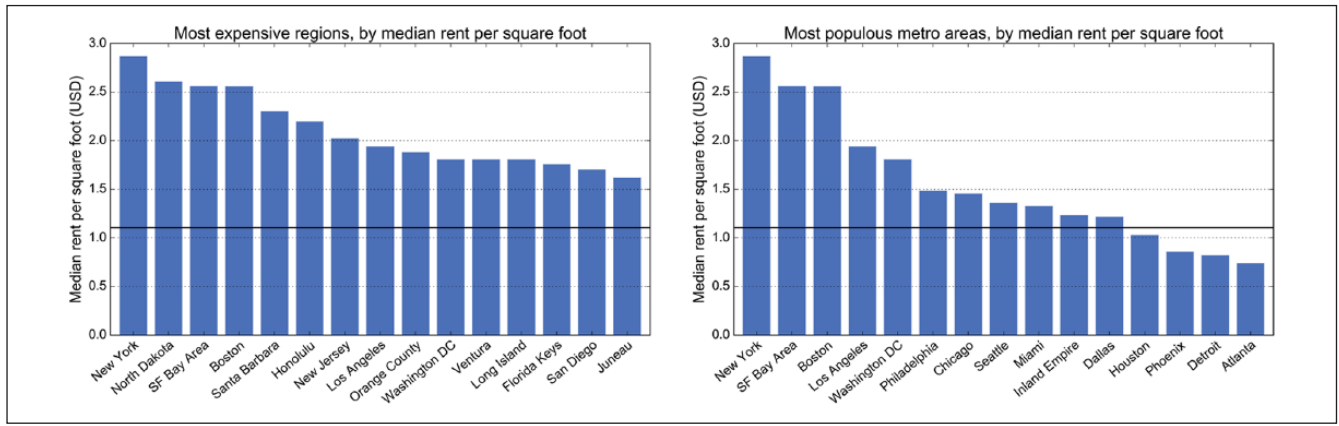

**Figure 2.** The most expensive (left) and most populous (right) Craigslist regions in the filtered data set, by median rent per square foot.
Note: Horizontal line depicts the nationwide median.

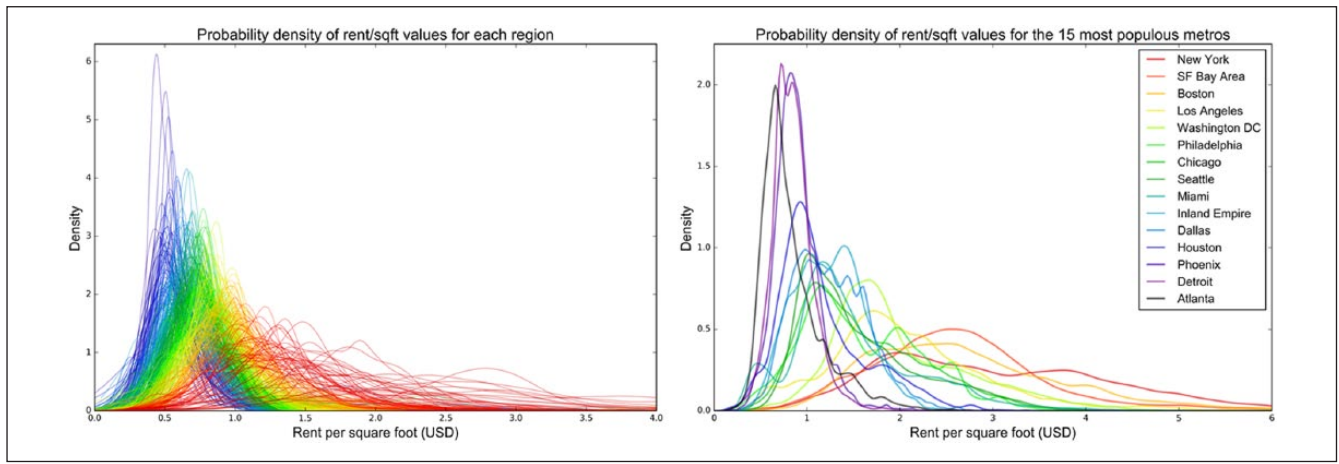

**Figure 3.** Probability densities of rent per square foot in the filtered data set for all 415 Craigslist regions (left), with the 15 most populous broken out for detail (right).
Note: Each region has its own line, colored by its median rent per square foot. The area under the curve between any two points represents the probability for that interval.

Different regions have different statistical distributions of rent per square foot values. Most are heavily right-tailed, but this is a function of the region's median value. We estimated the frequency of rent per square foot values for all 415 regions, each represented by its own line (Figure 3). These data show the heavy-tailed distributions that would be expected of diverse and heterogeneous spatial data (Anderson 2006). The regions with the most skewed distributions also indicate a more extreme gap between the highest end and the rest of the market. The gradient reveals the relationship between rent per square foot's per-region median, mode, and statistical dispersion: regions with lower median rents per square foot tend to peak at lower values and tend to be *more* peaked.

This compression of rents in soft markets is a significant finding for planners. In Detroit, most of the listed units are concentrated within a narrow band of rent per square foot values, but in San Francisco rents are much more dispersed. This suggests to practitioners and policymakers that FMR-based housing vouchers—designed to unlock neighborhoods of opportunity to the poor—may serve different functions in high-cost versus low-cost areas. The typical 40 percent FMR might be insufficient to upgrade neighborhoods in statistically dispersed markets like San Francisco, especially when considering household size.

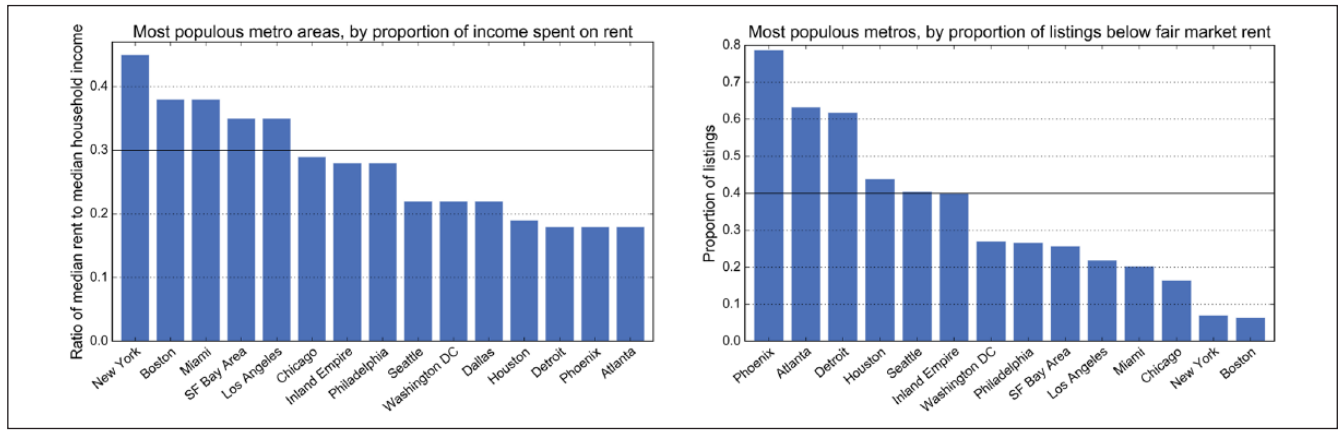

**Figure 4.** Left: Ratio of median rent to median monthly household income for the fifteen most populous metropolitan areas in the United States. Right: Proportion of listings in the filtered data set at/below HUD FMR.
Note: Horizontal lines depict the standard 30 percent definition of rent burden (left) and the standard 40th percentile FMR (right). FMR analysis excludes Dallas (see appendix). HUD = US Department of Housing and Urban Development; FMR = fair market rent.

### Affordability

"Rent burden" is typically defined by rent exceeding 30 percent of household income (Quigley and Raphael 2004; Schwartz 2010; Aratani et al. 2011). Although this flat ratio has been critiqued by some for oversimplifying affordability, it remains the standard convention in research and practice (USHUD 2014). We calculated the "rent proportion"—what share of its income a typical household would spend on a typical Craigslist rent—for each metropolitan area. At their *median* values, New York, Los Angeles, San Francisco, Miami, Boston, and San Diego all *exceed* the rent burden threshold. The most populous metropolitan areas' rent proportions demonstrate a wide variation in burden (Figure 4; details in appendix). This is a useful indicator of affordability for local and regional planners.

We also calculated the proportion of listings in the filtered data set at or below the HUD FMR, per region and number of bedrooms. HUD FMR values generally define the 40th percentile rent identified by regional surveys that exclude certain public and subsidized housing, among other requirements. We would expect these proportions of listings to typically be about 0.4, since the 40th percentile value is greater than 40 percent of all values. As previously discussed, the Craigslist data represent median advertised rents while the HUD data represent a sample of rents paid. Yet, in total, 37 percent of the listings in these regions are below the corresponding HUD FMR—quite close to the expected value of 40 percent. However, there is considerable variation. While more than two-thirds of the listings in regions like Phoenix, Las Vegas, and Kansas City are below the FMR, New York and Boston have only *single-digit* percentages of listings below this threshold (Figure 4; details in appendix). This is a troubling finding for planners. As discussed earlier, FMRs might be insufficient for households trying to upgrade neighborhoods in metros with highly dispersed rent values; they *also* appear to limit housing seekers in New York and Boston to very narrow slices of available housing units.

The disconnect between current listings and FMR levels might be due to an interaction of factors: (1) the prevalence of rent control in certain markets; (2) FMR calculations lagging behind the market; and (3) FMR calculations' basis on 5-year ACS estimates. For example, the HUD FMR for a two-bedroom unit in Alameda County, California (in the San Francisco Bay Area), *dropped* from $1,585 in 2015 to $1,580 in 2016—despite the region's skyrocketing rents since mid-2013—because HUD extrapolated FMR from the 2013 five-year ACS estimate. HUD currently requires housing authorities to conduct their own (time consuming and expensive) survey of rents to protest the established FMR. Craigslist data offer an invaluable real-time alternative to easily take the pulse of local housing rental markets at fine scales to inform superior, more-current estimates—particularly when subjected to some ground-truthing or supplemented with a limited traditional survey.

Intermetropolitan variation is ripe for future research, as it may offer nuance to HUD values or shed light on local market behaviors. Of the 15 most populous metropolitan areas in the United States, the large cities in California and along the Eastern Seaboard have the highest median rents per square foot, while those of large cities elsewhere in the Sunbelt (plus Detroit) are much lower (Figure 2). New York's median rent per square foot is more than 3.5 times higher than Atlanta's, reflecting underlying differences in land values that capitalize intermetropolitan variation in amenities, incomes, demand, and supply. Conversely, the "rental power" indicator represents an estimate of how many square feet can be rented in each region (given the median rent per square foot for each) for the nationwide median rent of $1,145.

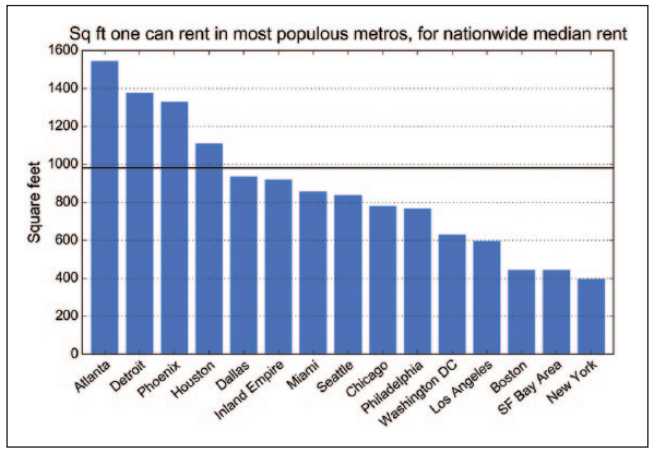

**Figure 5.** Rental power indicator: how many square feet one can rent in the fifteen most populous metropolitan areas, for the nationwide median rent.
Note: Horizontal line represents the nationwide median square footage in the filtered data set.

Memphis offers the greatest value among these regions at 1,659 square feet, while New York offers the least at 398 square feet (Figure 5; details in appendix). This indicator facilitates nationwide comparisons but does not explicitly represent what a typical worker can afford in each region, as wages vary between them.

### Metropolitan Spatial Patterns

Planning practitioners require current data in particular at the local scale. The Craigslist data set is perfectly suited for this. To demonstrate, we calculated tract-level median rents per square foot from one hundred thousand San Francisco Bay Area listings (Figure 6). This provides a snapshot of rental costs in the moment at the neighborhood scale. Real-time, fine-grained data are invaluable for tracking a hot housing market, and such visualizations enable local planners to quickly take the pulse of changing neighborhoods. Other useful indicators, such as rents per unit size/bedrooms, and at other spatial scales, such as the municipality or the state, are just as easily calculated on demand from these data.

Given the breadth and depth of this data set, analysts can examine rich spatial patterns within any metropolitan area, painting a portrait of the market for regional planners. Northern California exhibits a clear pattern of high rents near the coast and lower rents further inland. San Francisco, Berkeley, and Silicon Valley are very expensive, while Santa Rosa and Vallejo are slightly less so. Further inland, Sacramento tends to be midpriced, while rents in the smaller cities south of it through the Central Valley are in the lowest quintiles. Likewise, the greater Los Angeles area shows (in Figure 7) a clear gradient in rents from expensive coastal areas, like the west side of Los Angeles, towards cheaper inland areas in the East. Although some inland urban areas like San Bernardino and Palm Springs have moderate rents, others like Victorville and Hemet are heavily represented by the lowest quintiles.

In many ways, California's climate, geography, affluence, and land-use policies make it an anomaly (Fulton and Shigley 2012), and the Craigslist data reveal different spatial patterns elsewhere in the country. In the Midwest, Chicago has a high-priced urban core with midpriced suburbs, but Detroit inverts this metropolitan model with a low-priced core and more expensive exurbs. This is consistent with our expectations given Detroit's history of capital flight and white flight over the past fifty years (Sugrue 2005). Likewise, small "rust belt" cities in Indiana, Ohio, and Michigan are dominated by listings in the lowest quintiles.

### Usage Trends

In addition to these market patterns, we also examined the Craigslist data to better understand its usage trends. As discussed earlier, the raw Craigslist data obtained from web scraping contained duplicates, which we identified and filtered. Interestingly, the count and proportion of duplicates vary greatly by region. The listings posted in Seattle and Los Angeles are more likely to be unique and to provide complete data for the housing unit. In contrast, listings in Chicago and New York are more likely to have incomplete data and to be re-posted multiple times. Local practitioners thus may have "better" data in some markets than others.

There is a repeating cyclical pattern across the dates in the data set, as certain days of the week consistently have more listing activity than others. Mondays and Tuesdays have the greatest rental listing activity and Sundays the least. The periodic upward jumps in Figure 8 correspond to the weekly transition from Sunday to Monday in Table 3. From Monday onward, the number of daily listings declines, before repeating all over again. Housing practitioners thus might work with households to target housing searches to take advantage of these rhythms in volume, cost, and unit characteristics: the median rent per square foot drifts generally upward over the course of the week, in contrast to the general downward trend in the number of listings posted per day. Median rents per square foot are about 11.5 percent higher on Sundays (the most expensive day) than they are on Mondays (the least expensive day). It also appears that Mondays have a disproportionately high ratio of low-quality listings that were filtered out. In contrast, Tuesday's share of the total listings increased after cleaning and filtering the data set, indicating a disproportionately high ratio of quality listings posted on Tuesdays.

### Preliminary Validation

While listings are not the same as the final rents negotiated and formalized in a rental contract, they provide the closest available approximation of true rents. We might expect that listed rents are higher than final rents transacted in contracts as a result of negotiation if the average housing market is "soft" (more listings than actively searching renters). On the other hand, in extremely "tight" housing markets, competition for scarce rental housing might bid up rents above the listed value. This effect may also vary by the characteristics of the rental unit. A thorough investigation of this issue is beyond the scope of this present study, and we defer to future research the challenge of developing a method to adjust for it.

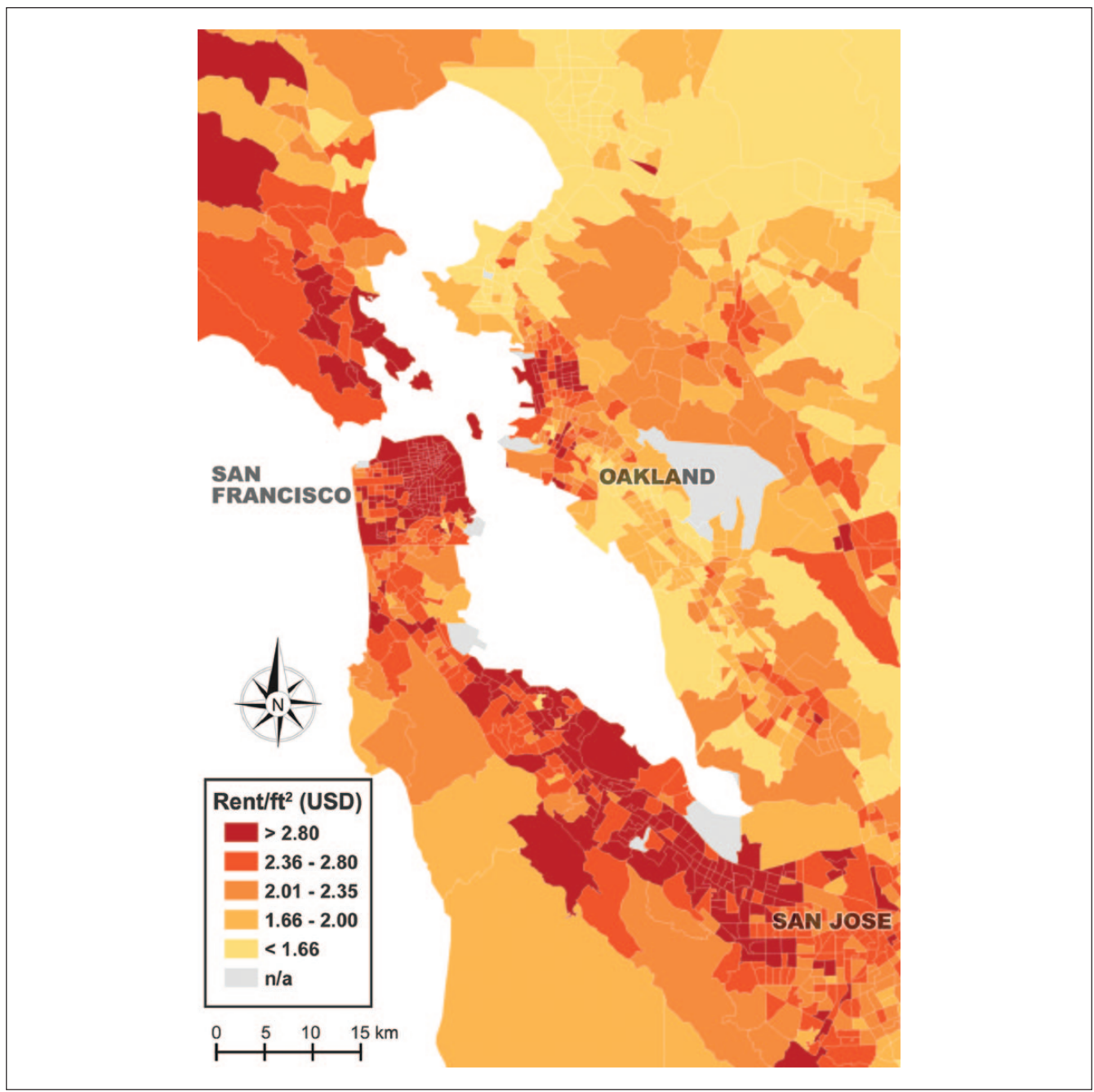

**Figure 6.** Census tract–level median rents per square foot in the San Francisco Bay Area.

However, we conducted a preliminary validation by comparing Craigslist data to HUD's estimated metropolitan area median rents (USHUD 2015). One may initially expect incomparability between HUD data and Craigslist rents given the caveats above and the fact that HUD data cover entire years. Nevertheless, and thus all the more interesting, our preliminary validation reveals surprising comparability. We divided the Craigslist filtered data set's median rent for each region and number of bedrooms by the HUD 2014 median rent for the corresponding metropolitan area and number of bedrooms. The result is the ratio of each Craigslist median rent to the corresponding HUD median rent. We then calculated correlation coefficients to assess the covariation between these data (see appendix). The two data sets are highly correlated but demonstrate heteroscedasticity: more expensive metropolitan areas are more dissimilar (Figure 9). The correlations between the Craigslist and HUD median rents are positive, strong, and statistically significant (Table 4).

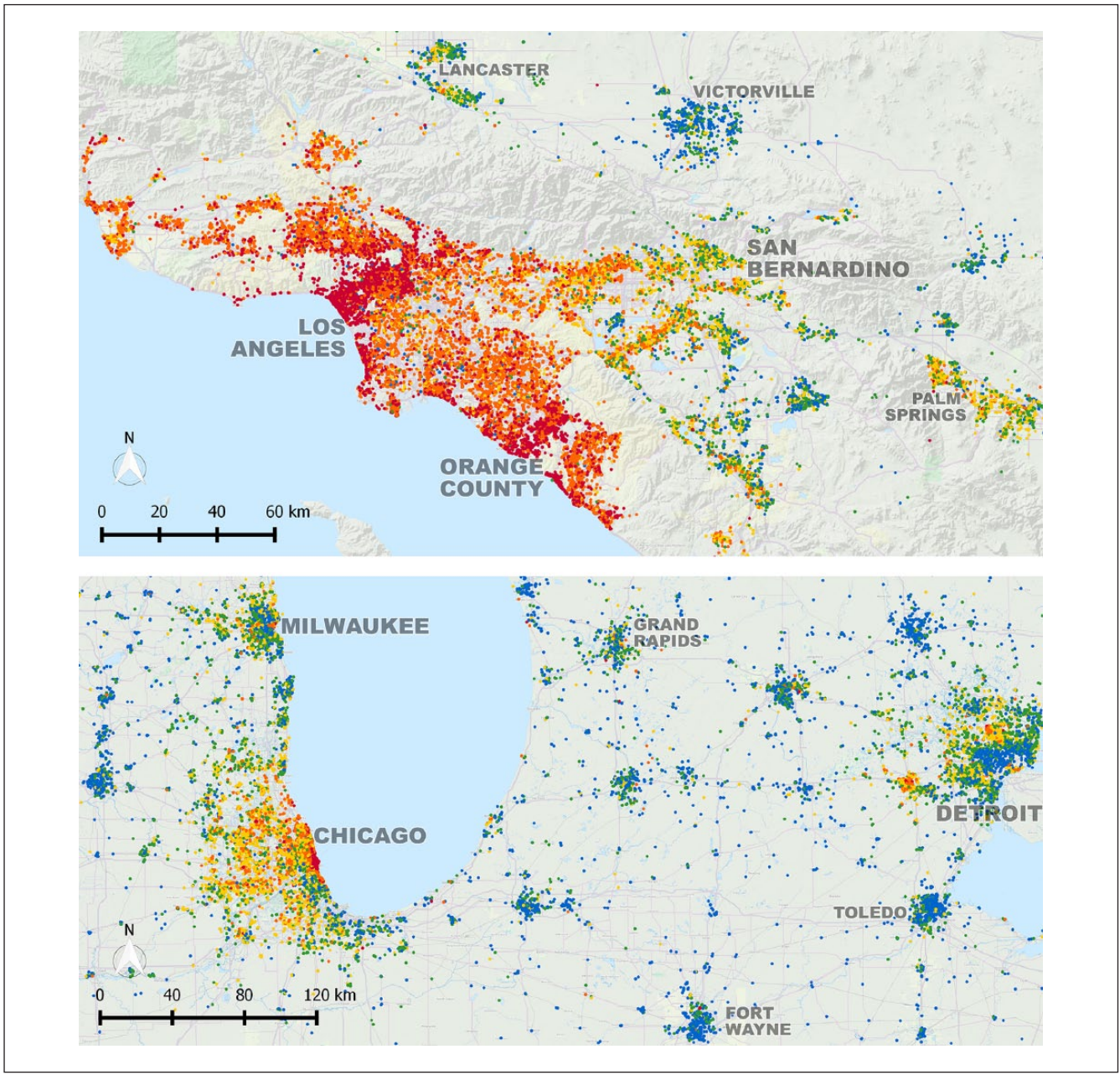

**Figure 7.** Detail of rental listings in the greater Los Angeles area (top) and the Midwest between Chicago and Detroit (bottom).
Note: Listings are divided into nationwide quintiles by rent per square foot (see Figure 1 legend).

To compare Craigslist and HUD metropolitan area median rents, we used the ratio of each Craigslist median rent to the corresponding HUD median rent (see appendix). This indicator also has the benefit of straightforward interpretation for planners: on average in these regions, median rents in the filtered data set are 7 percent higher for one-bedroom, 3 percent higher for two-bedroom, 7 percent lower for three-bedroom, and 1 percent higher for four-bedroom units than the corresponding HUD median rents. However, the bias varies between regions and numbers of bedrooms: New Orleans and Oklahoma City have very similar median rents across the two data sets, while other regions like Las Vegas have considerably *lower* median rents in the Craigslist data set, and yet other regions like New York have considerably *higher* median rents in the Craigslist data set.

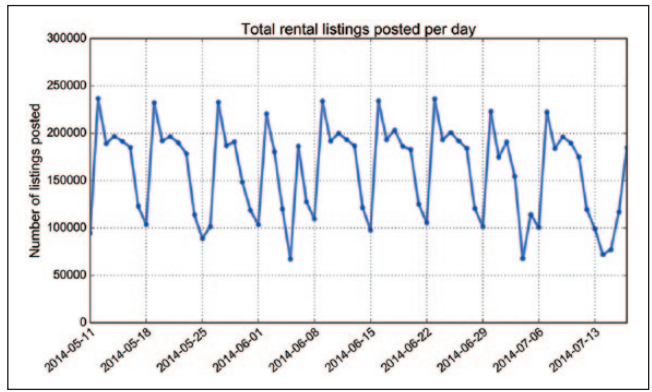

**Figure 8.** Count of rental listings posted on each date.
Note: Sundays are labeled.

**Table 3.** Median Rent per Square Foot and Mean Count of Listings Posted, by Day of the Week in the Filtered Data Set.

| Day of the Week | Median Rent/ft$^2$ | Mean Count |
|---|---|---|
| Monday | $1.09 | 50,848 |
| Tuesday | $1.09 | 52,256 |
| Wednesday | $1.08 | 49,836 |
| Thursday | $1.10 | 46,771 |
| Friday | $1.10 | 43,790 |
| Saturday | $1.17 | 33,188 |
| Sunday | $1.20 | 25,785 |

Craigslist may represent varying, limited segments of housing markets in different regions. Alternatively, market actors may behave differently in different regions, intentionally advertising rents above or below the anticipated final negotiated rent—behavioral differences that may also be correlated with market conditions. This is an interesting finding for future research to explore further. Census data are also affected by statistical and nonstatistical errors (such as nonresponse bias): differences between the two data sets could represent empirical differences or differences between unknown error terms. Nevertheless, given the concerns around small-sample five-year ACS estimates (Bazuin and Fraser 2013; Folch et al. 2014), the Craigslist data have something novel to tell planners: they offer generally reliable and up-to-date "small area" estimates. They provide richer detail down to the neighborhood scale on unit characteristics such as square footage and number of bedrooms. ACS tract-level median rent does *not* tell practitioners what a family of four needs to pay this month to lease a three-bedroom unit. The Craigslist data set *does*, across the United States.

## Prospects and Challenges for Urban Big Data

VGI, such as these Craigslist listings, provides enormous volumes of urban data to discover new questions to ask as

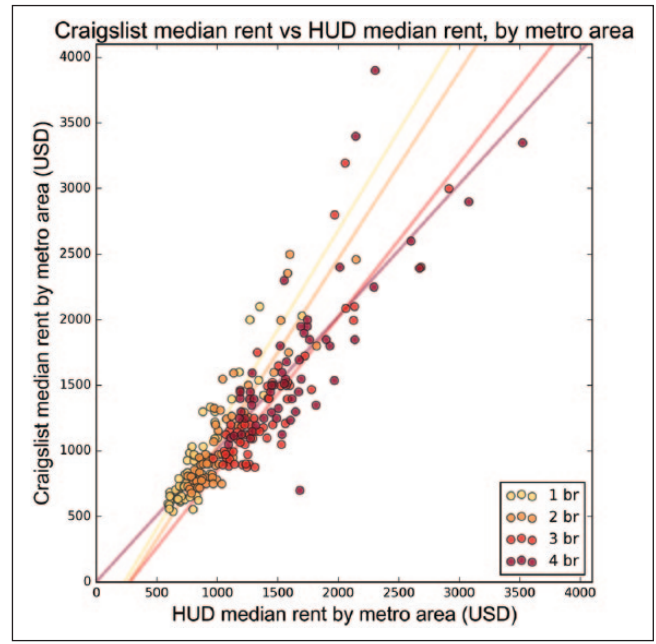

**Figure 9.** Scatterplot of Craigslist median rents versus HUD median rents for each sampled metropolitan area.
Note: Simple regression lines represent the relationship between the two data sets disambiguated by number of bedrooms. HUD = US Department of Housing and Urban Development.

**Table 4.** Correlational Analysis of the Craigslist Median Rents and the HUD Median Rents across Regions, by Number of Bedrooms.

| Bedrooms | r | Significance |
|---|---|---|
| 1 | .91 | $p < .0001$ |
| 2 | .90 | $p < .0001$ |
| 3 | .88 | $p < .0001$ |
| 4 | .79 | $p < .0001$ |

Note: The Pearson product–moment correlation coefficients (r) are positive, strong, and statistically significant. HUD = US Department of Housing and Urban Development.

we study and plan cities. Collecting, cleaning, and analyzing these data reveal national and regional rental housing market behaviors that planners can use to assess cost trends, housing characteristics, and affordability. The usage patterns suggest important geographic variations in how individuals collectively interact with this sector of the rental housing market, revealing rhythms that could not have been anticipated without digging deeply into these data. Surprisingly, metros like New York and Boston have only single-digit percentages of listings below the corresponding HUD FMR—raising new questions about affordability and vouchers based on FMRs. The rental power indicator offers a useful high-level snapshot of intermetropolitan rental market variation. The Craigslist data provide planners unprecedented real-time information, including unit characteristics, from neighborhood to national scales. But along with the benefits of big data come challenges. Conclusions drawn from VGI, like social media posts (e.g., Evans-Cowley and Griffin 2012; Schweitzer 2014) or online rental listings, must be contextualized in their settings: the geographic variability in usage reminds us that this is not a uniform data set collected from actual rental contracts. Usage may vary between metropolitan markets and individuals based on Internet access and technical capacity.

Returning to the 4 Vs of big data, the *variety* of these data—from numerous geographies and individuals—creates a nationwide tapestry of rental market information, and the listings' unstructured titles and descriptions might even enable new text/language analyses. The *velocity* may be both a strength and a weakness. We have access to daily data, but the harvesting requires automated tools such as this web scraper to run regularly—and possibly constantly if every single listing is to be collected. However, we believe a constant collection would overburden Craigslist's servers, and consequently prefer a less-frequent process instead. The *volume* and *veracity* are linked: big data can be unwieldy and messy. Collecting, cleaning, managing, and analyzing these Craigslist data required substantial work to produce something capable of market analysis. Fortunately, planners and researchers interested in using this methodology will have a simple, reusable, on-demand tool after initial setup. Lastly, VGI is only as good as the input. On Craigslist, rents and square footages are entered as minimally validated text, and geolocation is provided by users dropping a pin onto a map indicating the rental unit's location. As with any data set, if listings are misspecified or incorrectly geolocated, their errors persist in our analysis.

Nevertheless, the law of large numbers (i.e., volume and veracity) appears to hold. The filtered data set yields millions of rental listings, and the spatial pattern of these listings conforms to our expectations. This data set gives planners a new tool to assess local, regional, and interregional rental housing markets. Further, the methodology described here can be easily generalized to other spatial and urban research projects by planning agencies and scholars. Web scraping and related methods can harness vast swaths of user-generated data from the web and other digital documents. These methods of cleaning, analyzing, and visualizing urban big data can grapple with the overwhelming volume of data that cities and citizens now produce—data only useful if they can be corralled and comprehended.

However, many planners may lack the technical capacity to run Python code or other ad hoc software solutions. Fortunately, a couple of new paths forward are emerging. First, the growing "civic tech" community offers institutional capacity building services through organizations like Code for America. As mentioned earlier, planners must bring their knowledge and voice into planning technology and not just leave the future up to technologists now turning their attention to cities. There is also a small but growing faction of planners with the requisite technical skills and interest to do this work. The urban informatics and visualization master's course we teach at UC Berkeley cultivates these skills, and we encourage other planning programs to consider developing and offering similar coursework. Finally, metropolitan planning organizations or university research centers with greater technical capacity might run reports for local planners, or might band together to work directly with Craigslist. A centralized data clearinghouse or reporting dashboard would reduce overhead and redundancy while making valuable housing market data available to planners nationwide. In particular, we encourage researchers to make use of the Internet Archive or to reach out to Craigslist directly to explore data partnering opportunities.

Future research can explore longer time slices of these rental listings, as analyzing multiple years would provide planners insights into seasonal and large-scale longitudinal market trends at local, metropolitan, or national scales. It would also provide better comparability to validate against HUD annual data. We conducted a preliminary validation of these data against HUD median rents, but further investigation should be conducted to more accurately estimate the direction and degree of bias at different scales. This would make these data more useful for modeling and forecasting. Monthly visual rental market reports could help planners more quickly identify emerging housing challenges, and predictive analytics built atop these data could enable planners and consumers to anticipate quickly evolving rental markets. Lastly, future research can construct hedonic models of rents to study housing markets. Such work could be extended to studies of gentrification and displacement as the fine spatial and temporal scales of these data give planners real-time indicators of local rental housing trends.

## Conclusion

Rentals compose a significant portion of the US housing market, but much of this market activity has been little-understood because of its informal characteristics and historically minimal data trail. In this study, we collected, cleaned, analyzed, and visualized eleven million Craigslist rental listings. The spatial patterns at national, regional, and local scales largely conform to expectations of the spatial variations in rents and housing density, yet provide far fresher and finer-grained data.

We assessed affordability by calculating rent burdens and proportions of listings below HUD FMRs for 58 metropolitan areas and found that 37 percent of the listings in these regions are below the corresponding FMR—but surprisingly some metros like New York and Boston are only in the single-digit percentages, suggesting significant affordability challenges to local practitioners. Our preliminary validation of these data indicates the rents are generally similar to corresponding HUD estimates, and the values across metropolitan areas in these two data sets are strongly correlated. But, crucially, the Craigslist data offer planners *current* data, including unit characteristics, from neighborhood to national scales.

These data thus enable up-to-date, fine-grained examinations of any metropolitan area in the United States and offer the most comprehensive data source currently available to examine its rental housing market. The data are publicly available to city planners and urban scholars seeking to understand these markets at various scales, and this methodology is generalizable to other nontraditional sources of urban and spatial big data. Beyond the "smart cities" hype, VGI and big data are having a paradigm-shifting impact on social science research, and urban planning is particularly well poised to take advantage of these new insights. Web scraping and large-scale data science open up a new world of data for practitioners and scholars to understand housing markets, urban dynamics, and collective human behavior.

## Technical appendix

### Analytical Toolkit

This project was conducted by the UC Berkeley Urban Analytics Lab and is publicly available at http://github.com/ual/. We cleaned, analyzed, and visualized these data using Python and its *pandas*, *numpy*, and *matplotlib* libraries. Python is a standard programming language for data science because it is free, fast, multipurpose, and powerful. Python is open-source and there is no cost to use it or its innumerable libraries of prepackaged functionality. Although Python is an interpreted language, its libraries provide compiled extensions to perform extremely fast vectorized functions. Further, one can use the same syntax and grammar to build statistical models, cartographic maps, software such as web scrapers, and complete web sites. It has become a Swiss army knife of the computational science world. Accordingly, it has grown popular and powerful, with countless researchers and developers contributing libraries to extend Python's capabilities. Today it is an ideal introductory language for planners entering the data science realm. For more resources on how to build a web scraper with Python, see Herman (2012), Feng (2014), and Scrapy Community (2015).

### Data Filtering

As discussed, our ranges for "reasonable" values used realistic upper and lower bounds (presented in Table 2) to filter the data set. The outcomes of this filtering confirmed its practicality. The rent per square foot standard deviation dropped sharply from 125,085.11 to 0.86 after filtering out only the 0.79 percent of listings that were the greatest outliers, now indicating a far more sensible statistical dispersion of the data (presented in Table 1). The mean rent per square foot also dropped sharply from 98.92 to 1.39—yet still higher than the median of 1.11, indicating a positive skew to the distribution. This positive skew is expected and common in diverse and heterogeneous spatial data (Jiang and Thill 2015): rents and square footages cannot drop below zero but can be arbitrarily large, producing a nonnormal distribution. Nevertheless, the filtering process yielded nonrobust statistics much more in the range of the robust statistics, as was hoped for.

### Validation Method

We conducted a preliminary validation by comparing the Craigslist data to HUD's metropolitan area rent estimates (see USHUD 2015). Dependent samples *t*-tests are commonly used to determine if the means of two samples are significantly different from each other. However, two-sample *t*-tests require that a set of Gaussian conditions be met: each sample must be generated via simple random sampling and the sampling distribution should be normal—that is, symmetric, unskewed, and without major outliers. Our data do not meet these conditions, which is unsurprising as spatially heterogeneous big data commonly violate the assumptions of Gaussian statistics (Jiang 2015). Instead, we calculated correlation coefficients to assess the covariation between these data sets. These results are presented in Table 4 and Figure 9.

## Data Appendix

### Data Appendix A

This table presents Craigslist rental market summaries by census statistical areas, sampled by those regions that either are one of the 50 most populous MSAs or among the 50 regions with the most total listings posted. 2014 median income is from the 2014 ACS 1-year estimates of median household income (in 2014 inflation-adjusted US dollars). 2014 population estimates are from the 2014 ACS annual estimates of resident population (as of July 1). Median rent, square foot, and rent per square foot are calculated from the filtered data set. The rent proportion is the ratio of median rent to median monthly household income. Rental power is an estimate of how many square feet can be rented in each region for the nationwide median rent, and is calculated by dividing nationwide median rent by regional median rent per square foot.

## Appendix A

| Census Bureau Statistical Area Name | 2014 Median Income | 2014 Population Estimate | Median Rent | Median ft$^2$ | Median Rent/ft$^2$ | Rent Proportion | Rental Power (ft$^2$) |
|---|---|---|---|---|---|---|---|
| Albuquerque, NM MSA | $47,581 | 904,587 | $750 | 900 | $0.87 | 0.19 | 1,316 |
| Atlanta–Sandy Springs–Roswell, GA MSA | $56,166 | 5,614,323 | $849 | 1,119 | $0.74 | 0.18 | 1,547 |
| Austin–Round Rock, TX MSA | $63,603 | 1,943,299 | $1,144 | 925 | $1.25 | 0.22 | 916 |
| Baltimore–Columbia–Towson, MD MSA | $71,501 | 2,785,874 | $1,225 | 945 | $1.28 | 0.21 | 894 |
| Birmingham–Hoover, AL MSA | $47,046 | 1,143,772 | $849 | 1,132 | $0.75 | 0.22 | 1,526 |
| Boston–Cambridge–Newton, MA-NH MSA | $75,667 | 4,732,161 | $2,400 | 961 | $2.56 | 0.38 | 447 |
| Buffalo–Cheektowaga–Niagara Falls, NY MSA | $50,074 | 1,136,360 | $810 | 1,000 | $0.86 | 0.19 | 1,331 |
| Charlotte–Concord–Gastonia, NC-SC MSA | $53,549 | 2,380,314 | $926 | 1,054 | $0.89 | 0.21 | 1,286 |
| Chicago–Naperville–Elgin, IL-IN-WI MSA | $61,598 | 9,554,598 | $1,500 | 1,000 | $1.46 | 0.29 | 784 |
| Cincinnati, OH-KY-IN MSA | $55,729 | 2,149,449 | $750 | 980 | $0.80 | 0.16 | 1,431 |
| Cleveland–Elyria, OH MSA | $49,889 | 2,063,598 | $775 | 1,000 | $0.84 | 0.19 | 1,363 |
| Columbus, OH MSA | $56,371 | 1,994,536 | $800 | 977 | $0.85 | 0.17 | 1,347 |
| Dallas–Fort Worth–Arlington, TX MSA | $59,530 | 6,954,330 | $1,100 | 953 | $1.22 | 0.22 | 938 |
| Denver–Aurora–Lakewood, CO MSA | $66,870 | 2,754,258 | $1,240 | 900 | $1.34 | 0.22 | 854 |
| Detroit–Warren–Dearborn, MI MSA | $52,462 | 4,296,611 | $800 | 1,000 | $0.83 | 0.18 | 1,379 |
| Fresno, CA MSA | $43,423 | 965,974 | $875 | 980 | $0.87 | 0.24 | 1,316 |
| Grand Rapids–Wyoming, MI MSA | $54,372 | 1,027,703 | $804 | 1,000 | $0.82 | 0.18 | 1,396 |
| Hartford–West Hartford–East Hartford, CT MSA | $68,532 | 1,214,295 | $1,100 | 1,000 | $1.11 | 0.19 | 1,031 |
| Houston–The Woodlands–Sugar Land, TX MSA | $60,072 | 6,490,180 | $962 | 917 | $1.03 | 0.19 | 1,111 |
| Indianapolis–Carmel–Anderson, IN MSA | $52,268 | 1,971,274 | $750 | 980 | $0.78 | 0.17 | 1,467 |
| Jacksonville, FL MSA | $51,117 | 1,419,127 | $795 | 1,058 | $0.80 | 0.19 | 1,431 |
| Kansas City, MO-KS MSA | $56,994 | 2,071,133 | $750 | 960 | $0.80 | 0.16 | 1,431 |
| Las Vegas–Henderson–Paradise, NV MSA | $51,214 | 2,069,681 | $800 | 1,030 | $0.78 | 0.19 | 1,467 |
| Los Angeles–Long Beach–Anaheim, CA MSA | $60,514 | 13,262,220 | $1,760 | 950 | $1.91 | 0.35 | 599 |
| Louisville/Jefferson County, KY-IN MSA | $50,932 | 1,269,702 | $750 | 1,000 | $0.78 | 0.18 | 1,467 |
| Memphis, TN-MS-AR MSA | $45,844 | 1,343,230 | $800 | 1,120 | $0.69 | 0.21 | 1,659 |
| Miami–Fort Lauderdale–West Palm Beach, FL MSA | $48,458 | 5,929,819 | $1,550 | 1,137 | $1.33 | 0.38 | 860 |
| Milwaukee–Waukesha–West Allis, WI MSA | $53,164 | 1,572,245 | $900 | 1,000 | $0.93 | 0.20 | 1,231 |
| Minneapolis–St. Paul–Bloomington, MN-WI MSA | $69,111 | 3,495,176 | $1,200 | 1,024 | $1.08 | 0.21 | 1,060 |
| Nashville–Davidson–Murfreesboro–Franklin, TN MSA | $52,640 | 1,792,649 | $999 | 1,053 | $0.94 | 0.23 | 1,218 |
| New Haven–Milford, CT MSA | $60,391 | 861,277 | $1,235 | 1,000 | $1.25 | 0.25 | 916 |
| New Orleans–Metairie, LA MSA | $46,784 | 1,251,849 | $999 | 1,000 | $1.01 | 0.26 | 1,133 |
| New York–Newark–Jersey City, NY-NJ-PA MSA | $67,066 | 20,092,883 | $2,500 | 900 | $2.87 | 0.45 | 398 |
| Oklahoma City, OK MSA | $52,416 | 1,336,767 | $759 | 1,000 | $0.77 | 0.17 | 1,487 |
| Orlando–Kissimmee–Sanford, FL MSA | $48,270 | 2,321,418 | $975 | 1,049 | $0.96 | 0.24 | 1,192 |
| Philadelphia–Camden–Wilmington, PA-NJ-DE-MD MSA | $62,171 | 6,051,170 | $1,445 | 950 | $1.49 | 0.28 | 768 |
| Phoenix–Mesa–Scottsdale, AZ MSA | $53,365 | 4,489,109 | $800 | 930 | $0.86 | 0.18 | 1,331 |
| Pittsburgh, PA MSA | $52,293 | 2,355,968 | $900 | 1,000 | $1.00 | 0.21 | 1,145 |

*(continued)*

**Appendix A. (continued)**

| Census Bureau Statistical Area Name | 2014 Median Income | 2014 Population Estimate | Median Rent | Median ft$^2$ | Median Rent/ft$^2$ | Rent Proportion | Rental Power (ft$^2$) |
|---|---|---|---|---|---|---|---|
| Portland–Vancouver–Hillsboro, OR-WA MSA | $60,248 | 2,348,247 | $1,113 | 925 | $1.15 | 0.22 | 995 |
| Providence–Warwick, RI-MA MSA | $55,836 | 1,609,367 | $1,150 | 1,000 | $1.17 | 0.25 | 978 |
| Raleigh–Cary, NC MSA | $62,313 | 1,242,974 | $900 | 1,000 | $0.91 | 0.17 | 1,258 |
| Richmond, VA MSA | $60,936 | 1,260,029 | $905 | 950 | $0.97 | 0.18 | 1,180 |
| Riverside–San Bernardino–Ontario, CA MSA | $54,586 | 4,441,890 | $1,295 | 1,000 | $1.24 | 0.28 | 923 |
| Rochester, NY MSA | $51,086 | 1,083,393 | $940 | 1,100 | $0.93 | 0.22 | 1,231 |
| Sacramento–Roseville–Arden Arcade, CA MSA | $60,015 | 2,244,397 | $995 | 889 | $1.08 | 0.20 | 1,060 |
| Salt Lake City, UT MSA | $62,642 | 1,153,340 | $895 | 950 | $0.98 | 0.17 | 1,168 |
| San Antonio–New Braunfels, TX MSA | $52,689 | 2,328,652 | $840 | 900 | $0.97 | 0.19 | 1,180 |
| San Diego–Carlsbad, CA MSA | $66,192 | 3,263,431 | $1,635 | 945 | $1.71 | 0.30 | 669 |
| San Jose–San Francisco–Oakland, CA CSA | $80,600 | 8,607,423 | $2,323 | 910 | $2.56 | 0.35 | 447 |
| Seattle–Tacoma–Bellevue, WA MSA | $71,273 | 3,671,478 | $1,305 | 900 | $1.36 | 0.22 | 841 |
| St. Louis, MO-IL MSA | $55,535 | 2,806,207 | $775 | 900 | $0.86 | 0.17 | 1,331 |
| Tallahassee, FL MSA | $44,242 | 375,751 | $750 | 1,000 | $0.79 | 0.20 | 1,449 |
| Tampa–St. Petersburg–Clearwater, FL MSA | $46,876 | 2,915,582 | $900 | 1,021 | $0.94 | 0.23 | 1,218 |
| Tucson, AZ MSA | $45,856 | 1,004,516 | $680 | 850 | $0.83 | 0.18 | 1,379 |
| Tulsa, OK MSA | $50,740 | 969,224 | $700 | 970 | $0.73 | 0.17 | 1,568 |
| Urban Honolulu, HI MSA | $74,634 | 991,788 | $1,700 | 800 | $2.20 | 0.27 | 520 |
| Virginia Beach–Norfolk–Newport News, VA-NC MSA | $58,871 | 1,716,624 | $968 | 1,024 | $0.97 | 0.20 | 1,180 |
| Washington–Arlington–Alexandria, DC-VA-MD-WV MSA | $91,193 | 6,033,737 | $1,687 | 900 | $1.81 | 0.22 | 632 |

Note: We used the San Jose–San Francisco–Oakland, CA CSA to accurately represent the region covered by the San Francisco Bay Area on Craigslist, and we combined Craigslist's separate regions for Los Angeles and Orange County into one to accurately represent the area covered by the census's Los Angeles–Long Beach–Anaheim, CA MSA.

## Data Appendix B

This table presents the proportion of listings in the filtered data set at or below the US Department of Housing and Urban Development (HUD) fair market rents (FMR), per sampled region and number of bedrooms. FMRs generally correspond to 40th percentile rents and FMR areas generally correspond to metropolitan areas, but HUD uses a more complicated formula to determine percentiles and area boundaries in different circumstances (USHUD 2007). While regions like Phoenix, Las Vegas, and Kansas City have greater than two-thirds of their listings below the fair market rent, New York and Boston have only single-digit percentages of listings below the fair market rent.

**Appendix B**

| Census Bureau Statistical Area Name | Proportion of Listings at/below HUD FMR | | | | |
|---|---|---|---|---|---|
| | 1-Bedroom | 2-Bedroom | 3-Bedroom | 4-Bedroom | All (1–4-Bedroom) |
| Albuquerque, NM MSA | 0.71 | 0.58 | 0.70 | 0.62 | 0.65 |
| Atlanta–Sandy Springs–Roswell, GA MSA | 0.47 | 0.62 | 0.81 | 0.74 | 0.63 |
| Austin–Round Rock, TX MSA | 0.30 | 0.30 | 0.44 | 0.45 | 0.33 |
| Baltimore–Columbia–Towson, MD MSA | 0.41 | 0.50 | 0.56 | 0.41 | 0.48 |
| Birmingham–Hoover, AL MSA | 0.29 | 0.25 | 0.50 | 0.41 | 0.34 |
| Boston–Cambridge–Newton, MA-NH MSA | 0.05 | 0.07 | 0.08 | 0.05 | 0.06 |
| Buffalo–Cheektowaga–Niagara Falls, NY MSA | 0.30 | 0.34 | 0.48 | 0.49 | 0.37 |

*(continued)*

**Appendix B. (continued)**

| Census Bureau Statistical Area Name | Proportion of Listings at/below HUD FMR | | | | |
|---|---|---|---|---|---|
| | 1-Bedroom | 2-Bedroom | 3-Bedroom | 4-Bedroom | All (1–4-Bedroom) |
| Charlotte–Concord–Gastonia, NC-SC MSA | 0.23 | 0.30 | 0.51 | 0.56 | 0.34 |
| Chicago–Naperville–Elgin, IL-IN-WI MSA | 0.15 | 0.16 | 0.21 | 0.18 | 0.16 |
| Cincinnati, OH-KY-IN MSA | 0.40 | 0.49 | 0.68 | 0.36 | 0.50 |
| Cleveland–Elyria, OH MSA | 0.39 | 0.48 | 0.63 | 0.43 | 0.49 |
| Columbus, OH MSA | 0.33 | 0.46 | 0.54 | 0.37 | 0.43 |
| Dallas–Fort Worth–Arlington, TX MSA[a] | n/a | n/a | n/a | n/a | n/a |
| Denver–Aurora–Lakewood, CO MSA | 0.06 | 0.11 | 0.23 | 0.22 | 0.11 |
| Detroit–Warren–Dearborn, MI MSA | 0.54 | 0.58 | 0.73 | 0.53 | 0.62 |
| Fresno, CA MSA | 0.30 | 0.57 | 0.55 | 0.41 | 0.51 |
| Grand Rapids–Wyoming, MI MSA | 0.27 | 0.34 | 0.65 | 0.42 | 0.41 |
| Hartford–West Hartford–East Hartford, CT MSA | 0.47 | 0.48 | 0.70 | 0.56 | 0.52 |
| Houston–The Woodlands–Sugar Land, TX MSA | 0.35 | 0.46 | 0.60 | 0.51 | 0.44 |
| Indianapolis–Carmel–Anderson, IN MSA | 0.49 | 0.53 | 0.62 | 0.58 | 0.54 |
| Jacksonville, FL MSA | 0.74 | 0.72 | 0.84 | 0.77 | 0.76 |
| Kansas City, MO-KS MSA | 0.62 | 0.69 | 0.76 | 0.54 | 0.68 |
| Las Vegas–Henderson–Paradise, NV MSA | 0.86 | 0.89 | 0.94 | 0.82 | 0.89 |
| Los Angeles–Long Beach–Anaheim, CA MSA | 0.14 | 0.17 | 0.43 | 0.47 | 0.22 |
| Louisville/Jefferson County, KY-IN MSA | 0.27 | 0.44 | 0.61 | 0.38 | 0.43 |
| Memphis, TN-MS-AR MSA | 0.50 | 0.59 | 0.69 | 0.60 | 0.61 |
| Miami–Fort Lauderdale–West Palm Beach, FL MSA | 0.13 | 0.13 | 0.39 | 0.27 | 0.20 |
| Milwaukee–Waukesha–West Allis, WI MSA | 0.28 | 0.35 | 0.40 | 0.33 | 0.34 |
| Minneapolis–St. Paul–Bloomington, MN-WI MSA | 0.19 | 0.23 | 0.38 | 0.36 | 0.26 |
| Nashville–Davidson–Murfreesboro–Franklin, TN MSA | 0.27 | 0.29 | 0.46 | 0.29 | 0.33 |
| New Haven–Milford, CT MSA | 0.37 | 0.45 | 0.52 | 0.41 | 0.44 |
| New Orleans–Metairie, LA MSA | 0.44 | 0.47 | 0.45 | 0.42 | 0.46 |
| New York–Newark–Jersey City, NY-NJ-PA MSA | 0.07 | 0.05 | 0.11 | 0.06 | 0.07 |
| Oklahoma City, OK MSA | 0.45 | 0.52 | 0.48 | 0.33 | 0.48 |
| Orlando–Kissimmee–Sanford, FL MSA | 0.50 | 0.51 | 0.69 | 0.62 | 0.55 |
| Philadelphia–Camden–Wilmington, PA-NJ-DE-MD MSA | 0.19 | 0.22 | 0.48 | 0.33 | 0.27 |
| Phoenix–Mesa–Scottsdale, AZ MSA | 0.77 | 0.79 | 0.84 | 0.74 | 0.79 |
| Pittsburgh, PA MSA | 0.18 | 0.26 | 0.40 | 0.45 | 0.28 |
| Portland–Vancouver–Hillsboro, OR-WA MSA | 0.12 | 0.24 | 0.44 | 0.27 | 0.24 |
| Providence–Warwick, RI-MA MSA | 0.15 | 0.23 | 0.41 | 0.41 | 0.27 |
| Raleigh–Cary, NC MSA | 0.30 | 0.40 | 0.47 | 0.40 | 0.38 |
| Richmond, VA MSA | 0.48 | 0.65 | 0.69 | 0.56 | 0.61 |
| Riverside–San Bernardino–Ontario, CA MSA | 0.27 | 0.35 | 0.53 | 0.62 | 0.40 |
| Rochester, NY MSA | 0.35 | 0.38 | 0.34 | 0.27 | 0.35 |
| Sacramento–Roseville–Arden Arcade, CA MSA | 0.58 | 0.62 | 0.69 | 0.53 | 0.62 |
| Salt Lake City, UT MSA | 0.34 | 0.42 | 0.60 | 0.43 | 0.43 |
| San Antonio–New Braunfels, TX MSA | 0.39 | 0.40 | 0.56 | 0.33 | 0.42 |
| San Diego–Carlsbad, CA MSA | 0.17 | 0.24 | 0.39 | 0.36 | 0.25 |
| San Jose–San Francisco–Oakland, CA CSA | 0.18 | 0.26 | 0.37 | 0.47 | 0.26 |
| Seattle–Tacoma–Bellevue, WA MSA | 0.25 | 0.42 | 0.67 | 0.58 | 0.41 |
| St. Louis, MO-IL MSA | 0.50 | 0.56 | 0.57 | 0.40 | 0.54 |
| Tallahassee, FL MSA | 0.86 | 0.89 | 0.80 | 0.85 | 0.85 |
| Tampa–St. Petersburg–Clearwater, FL MSA | 0.52 | 0.60 | 0.64 | 0.66 | 0.59 |
| Tucson, AZ MSA | 0.77 | 0.78 | 0.81 | 0.73 | 0.78 |
| Tulsa, OK MSA | 0.52 | 0.62 | 0.57 | 0.27 | 0.55 |

*(continued)*

**Appendix B. (continued)**

| Census Bureau Statistical Area Name | Proportion of Listings at/below HUD FMR | | | | |
|---|---|---|---|---|---|
| | 1-Bedroom | 2-Bedroom | 3-Bedroom | 4-Bedroom | All (1–4-Bedroom) |
| Urban Honolulu, HI MSA | 0.45 | 0.51 | 0.63 | 0.56 | 0.52 |
| Virginia Beach–Norfolk–Newport News, VA-NC MSA | 0.75 | 0.79 | 0.86 | 0.76 | 0.80 |
| Washington–Arlington–Alexandria, DC-VA-MD-WV MSA | 0.21 | 0.25 | 0.48 | 0.55 | 0.27 |
| Total across all of these metros | 0.29 | 0.36 | 0.51 | 0.45 | 0.37 |

[a]Dallas is excluded from this particular analysis as the Dallas metropolitan FMR area uses only disaggregate "Small Area FMRs" defined by ZIP codes.

## Data Appendix C

This table presents the ratio of the Craigslist median rent to the corresponding HUD median rent (see USHUD 2015) in each sampled region. Each ratio is calculated by dividing the Craigslist filtered data set median rent (for that region and number of bedrooms) by the HUD 2014 50th percentile rent for the corresponding metropolitan area. On average across these regions, median rents in the Craigslist filtered data set are 7 percent higher for 1 bedroom, 3 percent higher for 2 bedrooms, 7 percent lower for 3 bedrooms, and 1 percent higher for 4 bedrooms than the corresponding HUD median rents.

**Appendix C**

| Census Bureau Statistical Area Name | Ratio of Craigslist Median Rent to HUD Median Rent | | | |
|---|---|---|---|---|
| | 1-Bedroom | 2-Bedroom | 3-Bedroom | 4-Bedroom |
| Albuquerque, NM MSA | 0.88 | 0.92 | 0.85 | 0.88 |
| Atlanta–Sandy Springs–Roswell, GA MSA | 0.97 | 0.86 | 0.71 | 0.73 |
| Austin–Round Rock, TX MSA | 1.09 | 1.16 | 1.05 | 1.05 |
| Baltimore–Columbia–Towson, MD MSA | 1.10 | 1.01 | 0.94 | 1.12 |
| Birmingham–Hoover, AL MSA | 1.04 | 1.06 | 0.92 | 0.94 |
| Boston–Cambridge–Newton, MA-NH MSA | 1.58 | 1.49 | 1.42 | 1.59 |
| Buffalo–Cheektowaga–Niagara Falls, NY MSA | 1.13 | 1.07 | 0.98 | 0.96 |
| Charlotte–Concord–Gastonia, NC-SC MSA | 1.12 | 1.07 | 0.93 | 0.89 |
| Chicago–Naperville–Elgin, IL-IN-WI MSA | 1.47 | 1.48 | 1.31 | 1.48 |
| Cincinnati, OH-KY-IN MSA | 1.01 | 0.94 | 0.85 | 1.09 |
| Cleveland–Elyria, OH MSA | 1.00 | 0.97 | 0.88 | 0.99 |
| Columbus, OH MSA | 1.07 | 0.98 | 0.91 | 1.14 |
| Dallas–Fort Worth–Arlington, TX MSA | 1.28 | 1.36 | 0.96 | 1.07 |
| Denver–Aurora–Lakewood, CO MSA | 1.31 | 1.28 | 1.10 | 1.15 |
| Detroit–Warren–Dearborn, MI MSA | 0.90 | 0.88 | 0.74 | 0.91 |
| Fresno, CA MSA | 1.04 | 0.90 | 0.91 | 1.01 |
| Grand Rapids–Wyoming, MI MSA | 1.02 | 1.03 | 0.81 | 1.02 |
| Hartford–West Hartford–East Hartford, CT MSA | 1.01 | 1.02 | 0.82 | 0.92 |
| Houston–The Woodlands–Sugar Land, TX MSA | 1.24 | 1.03 | 0.89 | 0.99 |
| Indianapolis–Carmel–Anderson, IN MSA | 0.95 | 0.92 | 0.85 | 0.89 |
| Jacksonville, FL MSA | 0.80 | 0.78 | 0.67 | 0.77 |
| Kansas City, MO-KS MSA | 0.86 | 0.83 | 0.70 | 0.90 |
| Las Vegas–Henderson–Paradise, NV MSA | 0.74 | 0.72 | 0.69 | 0.74 |
| Los Angeles–Long Beach–Anaheim, CA MSA | 1.36 | 1.31 | 1.01 | 0.98 |
| Louisville/Jefferson County, KY-IN MSA | 1.08 | 0.97 | 0.86 | 1.06 |
| Memphis, TN-MS-AR MSA | 0.93 | 0.86 | 0.79 | 0.86 |
| Miami–Fort Lauderdale–West Palm Beach, FL MSA | 1.33 | 1.20 | 1.00 | 1.19 |
| Milwaukee–Waukesha–West Allis, WI MSA | 1.07 | 1.04 | 1.05 | 1.18 |
| Minneapolis–St. Paul–Bloomington, MN-WI MSA | 1.15 | 1.15 | 0.99 | 1.01 |
| Nashville–Davidson–Murfreesboro–Franklin, TN MSA | 1.11 | 1.08 | 0.99 | 1.24 |

*(continued)*

**Appendix C. (continued)**

| Census Bureau Statistical Area Name | Ratio of Craigslist Median Rent to HUD Median Rent | | | |
|---|---|---|---|---|
| | 1-Bedroom | 2-Bedroom | 3-Bedroom | 4-Bedroom |
| New Haven–Milford, CT MSA | 1.12 | 1.06 | 0.98 | 1.15 |
| New Orleans–Metairie, LA MSA | 0.99 | 0.97 | 0.95 | 1.04 |
| New York–Newark–Jersey City, NY-NJ-PA MSA | 1.56 | 1.56 | 1.55 | 1.69 |
| Oklahoma City, OK MSA | 0.98 | 0.92 | 0.95 | 1.10 |
| Orlando–Kissimmee–Sanford, FL MSA | 0.95 | 0.95 | 0.86 | 0.87 |
| Philadelphia–Camden–Wilmington, PA-NJ-DE-MD MSA | 1.42 | 1.41 | 1.06 | 1.19 |
| Phoenix–Mesa–Scottsdale, AZ MSA | 0.84 | 0.83 | 0.78 | 0.79 |
| Pittsburgh, PA MSA | 1.16 | 1.12 | 1.05 | 0.98 |
| Portland–Vancouver–Hillsboro, OR-WA MSA | 1.18 | 1.10 | 0.98 | 1.11 |
| Providence–Warwick, RI-MA MSA | 1.21 | 1.22 | 0.98 | 1.02 |
| Raleigh–Cary, NC MSA | 1.04 | 1.00 | 0.98 | 1.04 |
| Richmond, VA MSA | 1.02 | 0.90 | 0.85 | 0.96 |
| Riverside–San Bernardino–Ontario, CA MSA | 1.17 | 1.13 | 0.96 | 0.93 |
| Rochester, NY MSA | 1.01 | 1.02 | 1.08 | 1.23 |
| Sacramento–Roseville–Arden Arcade, CA MSA | 0.93 | 0.90 | 0.88 | 0.97 |
| Salt Lake City, UT MSA | 1.02 | 0.97 | 0.87 | 0.98 |
| San Antonio–New Braunfels, TX MSA | 1.00 | 0.99 | 0.90 | 1.07 |
| San Diego–Carlsbad, CA MSA | 1.25 | 1.09 | 0.98 | 1.00 |
| San Jose–San Francisco–Oakland, CA CSA | 1.19 | 1.15 | 1.03 | 0.95 |
| Seattle–Tacoma–Bellevue, WA MSA | 1.24 | 1.01 | 0.83 | 0.87 |
| St. Louis, MO-IL MSA | 0.94 | 0.89 | 0.88 | 1.05 |
| Tallahassee, FL MSA | 0.70 | 0.78 | 0.73 | 0.42 |
| Tampa–St. Petersburg–Clearwater, FL MSA | 0.93 | 0.89 | 0.85 | 0.87 |
| Tucson, AZ MSA | 0.85 | 0.84 | 0.78 | 0.84 |
| Tulsa, OK MSA | 0.93 | 0.87 | 0.91 | 1.22 |
| Urban Honolulu, HI MSA | 1.03 | 0.99 | 0.89 | 0.94 |
| Virginia Beach–Norfolk–Newport News, VA-NC MSA | 0.88 | 0.84 | 0.77 | 0.78 |
| Washington–Arlington–Alexandria, DC-VA-MD-WV MSA | 1.15 | 1.10 | 0.94 | 0.90 |
| Arithmetic mean | 1.07 | 1.03 | 0.93 | 1.01 |

## Acknowledgments

The authors wish to thank Carolina Reid, Karen Frick, Karen Chapple, Jake Wegmann, Rocio Sanchez-Moyano, Nat Decker, Dara Adib, David Von Stroh, and three anonymous reviewers for their helpful comments and assistance.

## Declaration of Conflicting Interests

The author(s) declared no potential conflicts of interest with respect to the research, authorship, and/or publication of this article.

## Funding

The author(s) disclosed receipt of the following financial support for the research, authorship, and/or publication of this article: The authors wish to thank the MacArthur Foundation (grant 106476-0) for their support.

## Note

1. For full-color versions of these figures, please see the online edition of the journal.

## References

Alexa. 2015. "Craigslist.org Site Overview." Alexa—Actionable Analytics for the Web. http://www.alexa.com/siteinfo/craigslist.org.

Anderson, C. 2006. *The Long Tail*. New York: Hachette Books.

Aratani, Y., M. Chau, V. Wight, and S. Addy. 2011. "Rent Burden, Housing Subsidies and the Well-Being of Children and Youth." New York: National Center for Children in Poverty.

Batty, M., K. Axhausen, F. Giannotti, A. Pozdnoukhov, A. Bazzani, M. Wachowicz, G. Ouzounis, and Y. Portugali. 2012. "Smart Cities of the Future." *European Physical Journal Special Topics* 214 (1): 481–518. doi:10.1140/epjst/e2012-01703-3.

Bazuin, J., and J. Fraser. 2013. "How the ACS Gets It Wrong: The Story of the American Community Survey and a Small, Inner City Neighborhood." *Applied Geography* 45:292–302. doi:10.1016/j.apgeog.2013.08.013.

Belsky, E. 2013. *The Dream Lives on: The Future of Homeownership in America*. W13-1. Cambridge, MA: Joint Center for Housing Studies.

Bettencourt, L. 2013. "The Origins of Scaling in Cities." *Science* 340 (6139): 1438–41. doi:10.1126/science.1235823.

Bettencourt, L., and G. West. 2010. "A Unified Theory of Urban Living." *Nature* 467 (7318): 912–13.

Boeing, G. 2016. "Honolulu Rail Transit: International Lessons from Barcelona in Linking Form, Design, and Transportation." *Planext* 2:28–47. doi:10.17418/planext.2016.3vol.02.

Brown, C. 2013. "North Dakota Went Boom." *The New York Times Magazine*, January 31. http://www.nytimes.com/2013/02/03/magazine/north-dakota-went-boom.html.

Brown, K. 2014. "In the Rental Market, Craigslist May Be Undisruptable." *San Francisco Chronicle*, July 5. http://www.sfgate.com/technology/article/In-the-rental-market-Craigslist-may-be-5601847.php.

Cayo, M., and T. Talbot. 2003. "Positional Error in Automated Geocoding of Residential Addresses." *International Journal of Health Geographics* 2 (10): 1–12.

Ching, T.-Y., and J. Ferreira. 2015. "Smart Cities: Concepts, Perceptions and Lessons for Planners." In *Planning Support Systems and Smart Cities*, edited by S. Geertman, J. Ferreira, R. Goodspeed, and J. Stillwell, 145–68. Cham: Springer International.

Craigslist. 2015. "Craigslist Factsheet." Craigslist. https://www.craigslist.org/about/factsheet.

Craigslist Inc. v. 3Taps Inc. 2013, 942 F. Supp. 2d 962. United States District Court for the Northern District of California.

Decker, N. 2010. "Housing Discrimination and Craigslist." *The Current* 14 (1): 43–57.

Dowall, D. 1981. "Reducing the Cost Effects of Local Land Use Controls." *Journal of the American Planning Association* 47 (2): 145–53. doi:10.1080/01944368108977099.

Evans-Cowley, J., and G. Griffin. 2012. "Microparticipation with Social Media for Community Engagement in Transportation Planning." *Transportation Research Record* 2307:90–98. doi:10.3141/2307-10.

Feng, J. 2014. "A Week of Mining Seattle's Craigslist Apartment Pricing." *Jay's Life*, December 23. http://www.racketracer.com/2014/12/23/a-week-of-seattles-craigslist-apartment-pricing/.

Folch, D., D. Arribas-Bel, J. Koschinsky, and S. Spielman. 2014. "Uncertain Uncertainty: Spatial Variation in the Quality of American Community Survey Estimates." Working Paper 1, Arizona State University, Tempe.

Fulton, W., and P. Shigley. 2012. *Guide to California Planning*, 4th ed. Point Arena, CA: Solano Press.

Garde, A. 2015. "Affordable by Design? Inclusionary Housing Insights from Southern California." *Journal of Planning Education and Research* 0739456X15600033:1–16.

Goldman, E. 2013. "Craigslist Wins Routine but Troubling Online Trespass to Chattels Ruling in 3Taps Case." *Technology & Marketing Law Blog*, September 20. http://blog.ericgoldman.org/archives/2013/09/craigslist_wins_1.htm.

Goodspeed, R. 2015. "Smart Cities: Moving beyond Urban Cybernetics to Tackle Wicked Problems." *Cambridge Journal of Regions, Economy and Society* 8 (1): 79–92. doi:10.1093/cjres/rsu013.

Gordon, L. 2006. *Brokered Deception: The Hidden Perils of Online Real Estate Ads*. New York: The Council of the City of New York.

Hanson, A., and Z. Hawley. 2011. "Do Landlords Discriminate in the Rental Housing Market? Evidence from an Internet Field Experiment in US Cities." *Journal of Urban Economics* 70 (2): 99–114. doi:10.1016/j.jue.2011.02.003.

Hau, L. 2006. "Newspaper Killer." *Forbes*, December 11. http://www.forbes.com/2006/12/08/newspaper-classified-online-tech_cx-lh_1211craigslist.html.

Herman, M. 2012. "Scraping Web Pages with Scrapy." *Michael Herman*, November 5. http://mherman.org/blog/2012/11/05/scraping-web-pages-with-scrapy/.

Holeywell, R. 2011. "North Dakota's Oil Boom Is a Blessing and a Curse." *Governing*, August.

Jiang, B. 2015. "Geospatial Analysis Requires a Different Way of Thinking: The Problem of Spatial Heterogeneity." *GeoJournal* 80 (1): 1–13. doi:10.1007/s10708-014-9537-y.

Jiang, B., and J.-C. Thill. 2015. "Volunteered Geographic Information: Towards the Establishment of a New Paradigm." *Computers, Environment and Urban Systems* 53:1–3. doi:10.1016/j.compenvurbsys.2015.09.011.

Joint Center for Housing Studies. 2013. *America's Rental Housing: Evolving Markets and Needs*. Cambridge, MA: Harvard University.

Kay, M. 2008. *XSLT 2.0 and XPath 2.0 Programmer's Reference*, 4th ed. Indianapolis, IN: Wrox.

Kitchin, R., and G. McArdle. 2016. "What Makes Big Data, Big Data? Exploring the Ontological Characteristics of 26 Datasets." *Big Data & Society* 3 (1). doi:10.1177/2053951716631130.

Kroft, K., and D. Pope. 2014. "Does Online Search Crowd Out Traditional Search and Improve Matching Efficiency? Evidence from Craigslist." *Journal of Labor Economics* 32 (2): 259–303. doi:10.1086/673374.

Kurth, R. 2007. "Striking a Balance between Protecting Civil Rights and Free Speech on the Internet: The Fair Housing Act vs. the Communications Decency Act." *Cardozo Arts & Entertainment Law Journal* 25:805–36.

Laney, D. 2001. *3D Data Management: Controlling Data Volume, Velocity, and Variety*. Report 949. META Group.

Lutz, C. 2014. "Build and Let Live: 40 Years of Affordable Housing in Aspen." *Aspen Journalism*, May 27. http://aspenjournalism.org/2014/05/27/build-and-let-live-40-years-of-affordable-housing-in-aspen/.

Macdonald, H. 2006. "The American Community Survey: Warmer (More Current), but Fuzzier (Less Precise) Than the Decennial Census." *Journal of the American Planning Association* 72 (4): 491–503. doi:10.1080/01944360608976768.

Mallach, A. 2010. *Meeting the Challenge of Distressed Property Investors in America's Neighborhoods*. New York, NY: Local Initiatives Support Corporation.

Mayer-Schönberger, V., and K. Cukier. 2013. *Big Data: A Revolution That Will Transform How We Live, Work, and Think*. Boston, MA: Eamon Dolan/Houghton Mifflin Harcourt.

Mitchell, R. 2015. *Web Scraping with Python: Collecting Data from the Modern Web*. Sebastopol, CA: O'Reilly Media.

Mossberger, K., C. Tolbert, D. Bowen, and B. Jimenez. 2012. "Unraveling Different Barriers to Internet Use: Urban Residents and Neighborhood Effects." *Urban Affairs Review* 48 (6): 771–810.

Oliveri, R. 2010. "Discriminatory Housing Advertisements On-Line: Lessons from Craigslist." *Indiana Law Review* 43:1125–83.

O'Neil, C., and R. Schutt. 2013. *Doing Data Science*. Sebastopol, CA: O'Reilly Media.

Pollock, R. 2016. "Policy: Urban Physics." *Nature* 531 (7594): S64–S66. doi:10.1038/531S64a.

Quigley, J., and S. Raphael. 2004. "Is Housing Unaffordable? Why Isn't It More Affordable?" *Journal of Economic Perspectives* 18 (1): 191–214.

Rae, A. 2015. "Online Housing Search and the Geography of Submarkets." *Housing Studies* 30 (3): 453–72. doi:10.1080/02673037.2014.974142.

Reid, J. 2015. *HTML5 Programmer's Reference*. New York: Apress.

Scheyder, E. 2015. "Reality Hits North Dakota's Pricey Apartment Market; Rents Drop." Reuters, March 2. http://www.reuters.com/article/2015/03/02/us-north-dakota-apartments-idUSK-BN0LY0TR20150302.

Schwartz, A. 2010. *Housing Policy in the United States*, 2nd ed. New York: Routledge.

Schweitzer, L. 2014. "Planning and Social Media: A Case Study of Public Transit and Stigma on Twitter." *Journal of the American Planning Association* 80 (3): 218–38. doi:10.1080/01944363.2014.980439.

Scrapy Community. 2015. "Scrapy Documentation." Scrapy: A Fast and Powerful Scraping and Web Crawling Framework. http://scrapy.org/doc/.

Seamans, R., and F. Zhu. 2014. "Responses to Entry in Multi-Sided Markets: The Impact of Craigslist on Local Newspapers." *Management Science* 60 (2): 476–93. doi:10.1287/mnsc.2013.1785.

Spielman, S., D. Folch, and N. Nagle. 2014. "Patterns and Causes of Uncertainty in the American Community Survey." *Applied Geography* 46: 147–57. doi:10.1016/j.apgeog.2013.11.002.

Splichal, C. 2015. "Recent Development: Craigslist and the CFAA: The Untold Story." *Florida Law Review* 67 (5): 1845–60.

Sugrue, T. 2005. *The Origins of the Urban Crisis: Race and Inequality in Postwar Detroit*. Princeton, NJ: Princeton University Press.

Townsend, A. 2013. *Smart Cities: Big Data, Civic Hackers, and the Quest for a New Utopia*. New York: Norton.

USHUD (US Department of Housing and Urban Development). 2007. *Fair Market Rents for the Section 8 Housing Assistance Payments Program*. Washington, DC: Office of Policy Development & Research.

USHUD (US Department of Housing and Urban Development). 2014. *Rental Burdens: Rethinking Affordability Measures. PD&R Edge*, September.

USHUD (US Department of Housing and Urban Development). 2015. *50th Percentile Rent Estimates*. HUD User Data Sets. http://www.huduser.gov/portal/datasets/50per.html.

Watts, D. 2007. "A Twenty-First Century Science." *Nature* 445 (7127): 489–89.

Wegmann, J., and K. Chapple. 2012. *Understanding the Market for Secondary Units in the East Bay*. WP-2012-03. Berkeley, CA: Institute of Urban and Regional Development.

Wolfe, N. 2015. "Using the Computer Fraud and Abuse Act to Secure Public Data Exclusivity." *Northwestern Journal of Technology and Intellectual Property* 13 (3): 301–16.

Zandbergen, P. 2008. "A Comparison of Address Point, Parcel and Street Geocoding Techniques." *Computers, Environment and Urban Systems* 32 (3): 214–32. doi:10.1016/j.compenvurbsys.2007.11.006.

## Author Biographies

**Geoff Boeing** is a PhD Candidate in the Department of City and Regional Planning at the University of California, Berkeley. His research interests include urban form, complexity, and urban data science.

**Paul Waddell** is a professor in the Department of City and Regional Planning at the University of California, Berkeley. His research interests include land use and transportation planning, urban simulation, visualization, and urban data science.